\documentclass[twocolumn,8pt, showpacs, aps,  prd,  a4paper, superscriptaddress,  nofootinbib, tightenlines,  floats,floatfix]{revtex4}
\usepackage{graphicx}
\usepackage{epsfig}
\usepackage{caption}
\usepackage{latexsym}
\usepackage{amsmath,amssymb}
\usepackage{lipsum}
\usepackage{color}

\newcommand{\beq}{\begin{equation}}
\newcommand{\eeq}{\end{equation}}
\newcommand{\beqa}{\begin{eqnarray}}
\newcommand{\eeqa}{\end{eqnarray}}
\newcommand{\bea}{\begin{array}}
\newcommand{\ena}{\end{array}}

\def\be{\begin{equation}}
\def\ee{\end{equation}}
\def\bea{\begin{eqnarray}}
\def\eea{\end{eqnarray}}

\def\4pig{\sfrac{4\pi G}{c^{4}}}

\def\hsp5{\hspace{5mm}}
\newcommand{\sfrac}[2]{{\textstyle{#1\over#2}}}
\def\case#1/#2{\textstyle\frac{#1}{#2}}

\begin{document}
\title{ Holographic Entanglement Entropy  in 2D Holographic Superconductor via $AdS_3/CFT_2$}

\author         {Davood Momeni}
\affiliation    {Eurasian International Center for Theoretical Physics and Department of General \& Theoretical Physics, Eurasian National University, \\ Astana 010008, Kazakhstan}
\email{d.momeni@yahoo.com}

\author{Hossein Gholizade}
\affiliation{Department of Physics, Tampere University of Technology\\  P.O.Box 692, FI-33101 Tampere, Finland}
\email{hosein.gholizade@gmail.com}

\author{ Muhammad Raza}
\affiliation{Department of Mathematics, COMSATS Institute of Information Technology, Sahiwal 57000, Pakistan}
\affiliation{
State Key Lab of Modern Optical Instrumentation,Centre for Optical and Electromagnetic Research,
Department of Optical Engineering, Zhejiang University, Hangzhou 310058, China}
\email{mraza@zju.edu.cn}
\author         {Ratbay Myrzakulov}
\affiliation    {Eurasian International Center for Theoretical Physics and Department of General \& Theoretical Physics, Eurasian National University, \\ Astana 010008, Kazakhstan}
\email{rmyrzakulov@gmail.com}

\date{\today}
 \begin{abstract}
The aim of the present letter is to find  the holographic
entanglement entropy (HEE) in 2D holographic superconductors (HSC).
Indeed, it is possible to compute the exact form of this entropy due
to an advantage of approximate solutions  inside normal and
superconducting phases with backreactions. By making the UV and IR
limits applied to the integrals, an approximate expression for HEE
is obtained. In case the software cannot calculate minimal surface
integrals analytically, it offers the possibility to proceed with a
numerical evaluation of the corresponding terms.  We'll understand
how the area formula incorporates the structure of the domain wall
approximation.  We see that HEE  changes linearly with belt
angle. It's due to the extensivity of this type of entropy and the
emergent of an entropic force. We find that the wider belt angle
corresponds to a larger holographic surface.  Another remarkable
observation is that no "confinement/deconfinement" phase transition
point  exists in our $2D$ dual field theory. Furthermore we observe
that the slope of the HEE with respect to the temperature
$\frac{dS}{dT}$ decreases, thanks to the emergence extra degree of
freedom(s) in low temperature system. A first order phase transition
is detected near the critical point.

\end{abstract}

\pacs{11.25.Tq, 03.65.Ud,74.62.-c}
\maketitle

{\it Introduction} Our contemporary physical questions are appearing 
a bit harder.  Anti-de Sitter space/Conformal
Field Theory (AdS/CFT)  conjecture gives an abstract and still
largely conjectural approach which applies in very general
situations \cite{Maldacena}. It stated: weakly coupled gravitational
models at $AdS$ bulk are dual to a strongly coupled CFT on boundary.
This means that the strongly coupled quantum systems may correspond
precisely to black holes. Gauge/gravity duality is a frequent
application, particularly seen in those systems with strongly
coupling, like type II superconductors \cite{gubser}-\cite{Hart}.
The AdS/CFT movement seems particularly adept in its innovative
approach to reality. Its areas of research interest include
holographic superconductors, Quark-Gluon plasma, and
superconductor/superfluid in condensed matter physics, particularly
using qualitative approaches \cite{HerzogRev}-\cite{Momeni:2014efa}.
AdS/CFT has been used recently to produce a realistic model for
entanglement quantum systems \cite{hee1,hee2} (with conformal field
theory descriptions) with some success \cite{app1}-\cite{app14}, as
a geometric approach. In order to address this issues, we consider
two possible portions $\tilde{A}$(set A), $B=\tilde{A}'$ (the
complementary set) of a single quantum system upon which an Hilbert
space $\mathcal{H}_{\tilde{A}}\times \mathcal{H}_{\tilde{A}'}$ may
be based. We  consider the Von-Neumann entropy
$S_{X}=-Tr_{X}(\rho\log\rho)$ the best of the best for statistical
description, where $Tr$ is the quantum trace of quantum operator
$\rho$ over quantum basis $X$. If we compute $S_{\tilde{A}}$ and
$S_{\tilde{A}'}$, this is extremely useful  to see
$S_{\tilde{A}}=S_{\tilde{A}'}$. A further consequence, however, is
that Von-Neumann entropies are now more likely to identify it with a
region, the boundary of $\partial\tilde{A}$ \cite{Srednicki:1993im}.
More recently, studies on the role of analytical methods in
computation of the EE have been initiated
\cite{Peng:2014ira}-\cite{Romero-Bermudez:2015bma}. It must be
specially an outstanding note in the role of this type of entropy to
be specially computed to lower dimensional quantum systems as its
$AdS_3/CFT_2$ picture. Using a specially designed gravitational
dual, we use EE to explain our 2D dynamical phase transitions. We'll
investigate the reduced HEE of a strip geometry (belt) in three
dimensional AdS background. It can be calculated analytically in
terms of the cutoff length. The HEE and total length (angle) is
approximated by a function for which the minimal surface integral is
analytically solvable. After the normal form for the zero
temperature has been derived, the extent to which the criticality
regime $T\sim T_c$ may be solved analytically is covered. In case
the hand cannot calculate surface integrals analytically it offers
the possibility to proceed with a numerical evaluation of the
corresponding integrals. The variation in aggregation of the HEE
illustrates a magnifying ability to adapt different phase
transitions. At this point in the system the superconducting phase
is preferred to the normal phase, thereby presenting the  minimal
surface area.
\par
{\it Model for 2D HSC} The following action combines the accuracy of
AdS bulk modeling with the 2D quantum system on boundary
\cite{Liu:2011fy}-\cite{Nurmagambetov:2011yt}:
\begin{eqnarray}\label{action}
&&S=\int d^3
x\sqrt{-g}\Big[\frac{1}{2\kappa^2}(R+\frac{2}{L^2})\\&&\nonumber-\frac{1}{4}F^{ab}F_{ab}-|\nabla\phi-i
 A\phi |^2-m^2|\phi|^2\Big].
\end{eqnarray}
Here, $\kappa^2$ defines the three dimensional gravitational constant
$\kappa^2=8 \pi G_3$,  the Newton constant $G_3$, $L$ is the AdS radius, $m^2=m_{\phi}^2\in(-1,\infty)$ mass of scalar field, and $g=det(g_{\mu\nu})$.
For more accurate information we may also choose to fix a metric to AdS bulk  over a given range of coordinates:
\begin{eqnarray}\label{g}
ds^2=-f(r)e^{-\beta(r)}dt^2+\frac{d\gamma^2}{f(r)}+\frac{r^2}{L^2}dx^2~.
\end{eqnarray}
We may choose a temperature for CFT from our AdS black hole:
\begin{eqnarray}
&&T=\frac{f'(r_{+})e^{-\beta(r_+)/2}}{4\pi}.
\end{eqnarray}
We can adapt any conventional information  by substituting static functions for gauge field $A_{\mu}$ and scalar field $\phi$ for bulk:
\begin{eqnarray}
A_t=A(r)dt,\ \ \phi\equiv\phi(r).
\end{eqnarray}
We can also use static symmetry to adapt our metric to best showcase the normal state of our system in the absence of scalar field:
\begin{eqnarray}
f(r)=k+\frac{r^2}{L^2}-\kappa^2\mu^2\log r,\label{f0}\\
A(r)=\rho+\mu\log r.\label{Phi0}
\end{eqnarray}
where $k=-\frac{r_{+}^2}{L^2}+\kappa^2\mu^2\log r_{+}$, $\mu,\rho$
correspond to  the chemical potential and charge density in the dual
field theory respectively. Here, $r_+$ is the radius $r$ of the
event horizon $f(r_+)=0$ for a AdS black hole. The fields
$A_{\mu},\phi$ will satisfy regularity if they satisfy these
auxiliary boundary conditions:
\begin{eqnarray}\label{horizonboundry}
A(r_+)=0,\ \
\phi'(r_+)=\frac{m^2}{f'(r_+)}\phi(r_+),
\end{eqnarray}
and the metric ansatz satisfies:
\begin{eqnarray}\label{horizonmetric}
&&f'(r_+)=
\frac{2 r_+}{L^2}-2 \kappa ^2 r_+ \left[m^2
\phi (r_+)^2+\frac{1}{2}
e^{\beta (r_+)} A '(r_+)^2\right],\\&&
\beta '(r_+)=-4 \kappa ^2 r_+ \left[\frac{ A'
(r_+)^2 \phi (r_+)^2 e^{\beta
(r_+)}}{f'(r_+)^2}+\phi '(r_+)^2\right].
\end{eqnarray}
The AdS asymptotic expansions for the fields,  (\ref{f0},\ref{Phi0}), require the values of:
\begin{eqnarray}
&&\beta\rightarrow 0,\ \ f(r)\sim \frac{r^2}{L^2},\ \ A(r)\sim \mu \log r,\nonumber\\
&&\phi(r)\sim
\frac{<\mathcal{O}_->}{r^{\Delta-}}+\frac{<\mathcal{O}_+>}{r^{\Delta+}},\ \mbox{as } r\to \infty.
\end{eqnarray}
The use of $\Delta_\pm$  can  denote  conformal dimensions
$\Delta_\pm=1\pm\sqrt{1+m^2}$ . Let $<\mathcal{O}_{\pm}>$ denote the
standard vacuum expectation values (VEV) of dual operators
$\mathcal{O}_{\pm}$ in CFT.  A change of variable $z =
\frac{r_+}{r}$ has been applied, which essentially simplifies the
forms of the equations of motion:
\begin{eqnarray}
&&\phi ''+\frac{\phi '}{z}
\left[1+\frac{zf'}{f}-\frac{z\beta
'}{2}\right]+\frac{r_{+}^2\phi}{z^4} \left[\frac{
A^2 e^{\beta
}}{f^2}-\frac{m^2}{f}\right]=0~,\label{phiz}\\&&
A ''+\frac{A '}{z} \left[1-\frac{z\beta '}{2}\right]-\frac{2 r_{+}^2 A \phi ^2}{z^4f}=0~,\label{Az}\\&&
\beta '-  \frac{4 \kappa ^2r_{+}^2}{z^3} \left[\frac{A
^2 \phi^2 e^{\beta}}{f^2}-
\frac{z^4\phi
'^2}{r_{+}^2}\right]=0\label{betaz},
\\&&
f'-\frac{2 r_{+}^2}{L^2z^3}- \kappa ^2 z
e^{\beta} A'^2-\frac{2 \kappa ^2 m^2 r_{+}^2\phi ^2}{z^3}
\\&&\nonumber-\frac{2 \kappa ^2 r_{+}^2}{z^3} \left[\frac{ A
^2 \phi^2 e^{\beta}}{f}+\frac{f \phi
'^2z^4}{r_{+}^2}\right]=0,\label{fz}
\end{eqnarray}
Solving the above equation with $\phi\neq0, T<T_c$ is the principal purpose of the superconductivity program \cite{Liu:2011fy,Momeni:2013waa}. \par
{\it HEE proposal :} Following to the  proposal \cite{hee1,hee2},
suppose a field theory in $(d)D$  has a gravitational dual embedded
in $AdS_{d+1}$ bulk. The holographic algorithm is then used to
compute  the entanglement entropy of a region of space $\tilde A$
and its complement from the $AdS_{d+1}$ geometry of bulk:
\begin{eqnarray}
S_{\tilde A}\equiv S_{HEE} = \frac{Area(\gamma_{\tilde A})}{4G_{d+1}},
\end{eqnarray}
 We first compute the minimal $(d-1)$D mini-super surface $\gamma_{\tilde A}$. It had been
proposed to extend $\gamma_{\tilde A}|_{AdS_{d+1}}$ to bulk, but
with criteria to keep surfaces with same boundary  $\partial
\gamma_{\tilde A}$  and $\partial {\tilde A}$. The equal boundary is
the leading technique working to compute HEE via AdS/CFT .
\par
Several options for the parametrization of the $\tilde A$  are
available as well as different choices for the $\partial
\gamma_{\tilde A}$ in the bulk. We discuss the possibility of
computing HEE of 2D  systems using parametric representation in one
degree of freedom $\tilde A:=\{t=t_0,-\theta_0\leq
\theta\leq\theta_0,r=r(\theta)\}$. Minimization  can  be used on
Lagrangian which has a simple Beltrami form:
\begin{eqnarray}
\mathcal{L}\equiv\sqrt{r^2+\frac{r'(\theta)^2}{f(r)}}
\end{eqnarray}
Using the Beltrami identity, since
 $\partial_{\theta}\mathcal{L}=0$, computations were taken by using a  constant quantity $\mathcal{L}-r'\partial_{r'}\mathcal{L}=C$ designed for $\mathcal{L}$\footnote{We can call it "Energy".}.
\par
In this case, we define half of the total length (angle) $\theta_0$  and HEE in the more conventional forms, using the shorter list enumerated here:
\begin{eqnarray}
\label{theta0}&&\theta_0=\int_{0}^{\theta_0}{d\theta}=\int_{0}^{\theta_0}{\frac{C dr}{r\sqrt{f(r)(r^2-C^2)}}}\\&&
\label{shee}S_{HEE}\equiv \frac{1}{2G_3}\int_{0}^{\theta_0}{\frac{rdr}{\sqrt{f(r)(r^2-C^2)}}}
\end{eqnarray}
The aim of this letter is to evaluate the sensitivity of
(\ref{theta0},\ref{shee}) in the bulk of acute regimes of
temperature.\par
 {\it Sharp domain wall approximation:}
Studies are currently underway to develop and evaluate
(\ref{theta0},\ref{shee}) using numerical algorithms. Our aim is to
evaluate the (\ref{theta0},\ref{shee}) as an analytic tool. The
(\ref{theta0},\ref{shee})  are determined from the domain wall
approximation analysis \cite{Albash2012}. Domain wall idea is
proposed to investigate some aspects of the HEE along renormalization
group (RG) trajectories. The RG flow is defined as the $N=1$ SUSY
deformation of $N=4$ SUSY-YM theory. The geometry (metric) which we
will use is called  here domain wall geometry. These are Riemannian
$3D$ spaces which are assumed to be asymptotically AdS. The RG is a
flow from one dual geometry in the UV to another in the IR. These
regions are separated by an intermediate border, which can be
realized as a domain wall, which is connecting the two regions. The
position of such domain wall and its thickness are functions of the
dual field theory parameters like dual charge density $\rho$ or dual
chemical potential $\mu$. What we want to understand is how the HEE
incorporates the structure of the domain wall. Furthermore we want
to know which kind of the field theory quantities  are encoded in
domain wall parameters. In our $CFT_2$ case the RG flow is $(1+1)D$
and we suppose that the bulk geometry $AdS_3$ is idealized by a
sharp domain wall medium. The domain wall is seperating two $AdS_3$
regions via two different values for the cosmological constant. What
we want to obtain is the form of HEE (\ref{shee}) and belt angle
(\ref{theta0}) incorporate the structure of the domain wall. We
consider the $AdS_3$ is relating to RG flows in $(1+1)D$. For the
AdS radius in two regions, we suppose that $L_{IR}>L_{UV}$.
Furthermore we suppose that the length scale of AdS $L$, is defined
as the following:

\begin{equation}
L = \left\{ \begin{array}{lr}
L_{UV} \ , & r > r_{DW} \\
L_{IR}\ , & r < r_{DW}
\end{array} \right. \ .
\end{equation}
Suppose we have a sharp phase transition between two patches of the
AdS space time. We will try to locate it at  $r=r_{DW}$ . Here
$r_{DW}$ defines the position of the domain wall in the AdS radial
direction. Indeed, in the massless limit, $m^2=0$,  there exists an
intermediate radius $-\infty<r_m<0$ such that $\phi'(r_m)=0$. The
ideal candidate for $r_{DW}$ should be  $r_{DW}=r_m$. We always
assume that $r_{DW}<0$. So we assume that the following form is a
perfectly good tool for analytical evaluation:
\begin{equation}\label{Beltrami_const}
C\equiv \frac{r^2}{\sqrt{r^2+ \frac{r'(\theta)^2}{f(r)}}} = \left\{ \begin{array}{lr}
L_{UV} \ , & r > r_{DW} \\
{L_{\rm IR}} \ , & r < r_{DW}
\end{array} \right. \ .
\end{equation}
In the previous equation we took into account two  different AdS radii, $L_{UV} $ and $L_{IR}$ in each region.
 With the previous considerations, the equations (\ref{theta0},\ref{shee}) are easily integrated,
\begin{eqnarray}\label{l2}
\int_0^{\theta_0}d\theta=\theta_0=\theta_{\text{IR}}+\theta_{\text{UV}},
\end{eqnarray}
\begin{eqnarray}\label{IIR}
&&\theta_{\text{IR}}=\int_{r_*}^{r_{DW}}{\frac{L_{IR}dr}{r\sqrt{f_{IR}}\sqrt{r^2-L_{IR}^2}}},\label{thetaIR}\\&&  \theta_{\text{UV}}=\int_{r_{DW}}^{r_{UV}}{\frac{Ldr}{r\sqrt{f_{UV}}\sqrt{r^2-L^2}}}\label{thetaUV}
\end{eqnarray}
 here $r_*$ denotes  the ``turning'' point of the minimal surface $\gamma_{\tilde A}$. It is defined by $r'(\theta)|_{r=r^{*}}=0$. So we can choose to engage turning point with $r_{+}$ or $C$. We replaced the integrating out to  $r=+\infty$ by integrating out to large positive radius $r_{UV}$.  Indeed,  we assume that $r_{UV}$
stands out for  UV cutoff
\cite{Romero-Bermudez:2015bma}.
We will suppose that $r_*<r_{DW}$. It means that  the minimal surface drift onto the IR region. We can rewrite HEE as we like:

\begin{eqnarray}
&&S_{HEE}=\frac{1}{2G_3}\Big[S_{IR}+S_{UV}\Big],\\&&
\label{SIR}S_{IR}=\int_{r_*}^{r_{DW}}{\frac{rdr}{\sqrt{f_{IR}(r^2-L_{IR}^2)}}},\\&&\label{SUV}
S_{UV}=\int_{r_{DW}}^{r_{UV}}{\frac{rdr}{\sqrt{f_{UV}(r^2-L_{UV}^2)}}}.
\end{eqnarray}
In both cases IR,UV, the geometry of AdS has imposed tight constraints on the metric:

\begin{equation}\label{hIR}
f(r)\to f_{\text{IR}}=1,\ \mbox{as } r\to-\infty.
\end{equation}
\begin{equation}\label{hUV}
f(r)\to f_{UV}=\frac{r^2}{L^2},\ \mbox{as } r\to\infty,
\end{equation}
\par
Perhaps we'll compute the $\theta_{\text{IR}},\theta_{\text{UV}}$ for another purpose:
\begin{eqnarray}
&&\theta_{IR}=i\log\Big[\frac{r_{*}}{r_{DW}}\frac{\sqrt{L_{IR}^2-r_{DW}^2}-L_{IR}}{\sqrt{L_{IR}^2-r_{*}^2}-L_{IR}}\Big],\\&&
\theta_{UV}=\frac{\sqrt{r_{UV}^2-L^2}}{r_{UV}}-\frac{\sqrt{r_{DW}^2-L^2}}{r_{DW}}.
\end{eqnarray}
The entangelement entropy can be computed as the following:
\begin{eqnarray}
&&S_{IR}=\sqrt{r_{DW}^2-L_{IR}^2}-\sqrt{r_{*}^2-L_{IR}^2}
\\&&S_{UV}=\frac{iL^2}{L_{UV}}\log\Big[\frac{r_{DW}}{r_{UV}}\frac{\sqrt{L_{UV}^2-r_{UV}^2}-L_{UV}}{\sqrt{L_{UV}^2-r_{DW}^2}-L_{UV}}\Big]
\end{eqnarray}
We consider two regimes, the IR and UV:

{\bf UV limit:} we first consider $r_*>r_{DW}$. This showed that
$\gamma_{\tilde A}$ were already embedding deeply into the $AdS_3$
with boundary $r\to\infty$. To get a rough approximation of how much
entropy in the $\gamma_{\tilde A}$ would be in UV limit, we simplify
the problem by puting $\theta_{\text{IR}}=S_{\text{IR}}=0$ and
identifying $r_{DW}=r_*$ in  Eqs. (\ref{thetaUV}) and (\ref{SIR}):
\begin{eqnarray}
&&\theta_{UV}\sim-\frac{1}{2}(\frac{L}{r_{UV}})^2,\ \mbox{as } r_{UV}\to \infty\\&&
S_{UV}\sim(\pi\mp\frac{\pi}{2})\frac{ L^2}{L_{UV}}  ,\ \mbox{as } r_{UV}\to \infty
\end{eqnarray}
The second term  is written to indicate the presence of black hole (BH) area entropy. The first term indicate clearly that the classical BH entropy is reduced through quantum effects.

{\bf IR limit:} In case $r_*\ll r_{DW}$, i.e. the $\gamma_{\tilde A}$ extends deeply into the IR region. From Eqs. (\ref{thetaIR}), (\ref{thetaUV}),  (\ref{SIR}) and (\ref{SUV})  we obtain:
\begin{eqnarray}
&&\frac{\theta_0}{2}=-\frac{\pi}{2}-\frac{\sqrt{r_{UV}^2-L^2}-r_{UV}}{r_{UV}},\
\mbox{as } r_{*}\to-\infty,\\&&
\lim_{r_{*}\to-\infty}S_{IR}^{\mbox{finite }}=r_{DW},\\&&
S_{UV}=\frac{L^2}{r_{DW}}\\&&\nonumber+\frac{iL^2}{L_{UV}^2}\log
\left( {\frac {iL_{{{\it UV}}}+\sqrt {{r_{{{\it UV}}}}^{2}-{L_{{{
\it UV}}}}^{2}}}{r_{{{\it UV}}}}}. \right)
\end{eqnarray}
Perhaps not surprisingly, $S^{HEE,IR}_{\mbox{finite
}}=\frac{r_{DW}}{2G_3}$  is exactly the result we would have
computed if we were purely in the IR theory. It's interesting to
note that IR limit can't follow the similar UV form. However
according to the note above the BH area term calculated in IR limit
is $S_{HEE}=S_{BH}=\frac{r_{DW}}{2G_3}+\frac{L^2}{2G_3r_{DW}}$, so
there is a difference for this regime. Note that there were major
differences in the BH term in both regimes.\par We mention here
that because $L_{IR}>L_{UV}$, so we conclude that the dominated part
of HEE is the one which is calculated in the IR limit given by
$S_{IR}\sim \frac{L^2}{r_{DW}}\sim L_{IR}$, however $S_{UV}\sim
L_{UV}<S_{IR}$ which is obviously less than $S_{IR}$.\footnote{The
point is that these geometries are meant to represent an RG flow.
For high energies (the UV of the theory), the background looks like
planar AdS (and hence the $N=4$ SYM theory), whereas at low energies
(the IR of the theory), the background looks like a different AdS
(the low-energy limit of the field theory). The domain wall
approximation was meant to be a toy model of the superconducting
backgrounds  \cite{Tameem Albash}. }
\par
{\it HEE close to the  $T \lesssim T_c$ in the absence of scalar field $\phi(z)=0$:}.  The normal  phase can even be achieved from a list of functions:
\begin{eqnarray}
&&\phi_0= \beta_0=0,\ \ A_0=-\mu_c\log z,\\&&
f_0=\frac{r_{+c}^2}{L^2}(z^{-2}-1)+\kappa^2\mu_c^2\log z.
\end{eqnarray}
The EE between $\tilde{A}$ and its complement is given by:
\begin{eqnarray}
&&s_{\tilde{A}}=4 G_3 S_{HEE}=
2r_{*}^{-1}\int_{r_{UV}}^{r_{*}}{\frac{rdr}{\sqrt{f(r)(r^2-r_{*}^2)}}}
\end{eqnarray}
The technique of this computation was a rewrite of $s_{\tilde{A}}$
with better coordinate $z$:
\begin{eqnarray}
s_{\tilde{A}}=2r_{+}r_{*}\int_{z_{UV}}^{z_{*}}\frac{dz}{z^3\sqrt{f(z)}\sqrt{z^{-2}-z_{*}^{-2}}}.
\end{eqnarray}
The vicinity of the critical point  $T \lesssim T_c$ maybe served as a place for an equivalent form of integral:
\begin{eqnarray}
s_{\tilde{A}}=2r_{+c}r_{*}\int_{z_{UV}}^{z_{*}}\frac{dz}{z^3\sqrt{f_0}\sqrt{z^{-2}-z_{*}^{-2}}}\label{s-Tc}.
\end{eqnarray}
and
\begin{eqnarray}
&&\frac{\theta_0}{2}=r_{*}\int_{r_{UV}}^{r_{*}}\frac{dr}{r\sqrt{f(r)(r^2-r_{*}^2)}}\\&&\nonumber=
\frac{r_{*}}{r_{+}}\int_{z_{UV}}^{z_{*}}\frac{dz}{z\sqrt{f(z)(z^{-2}-z_{*}^{-2})}}.
\end{eqnarray}
At criticality  $T \lesssim T_c$:
\begin{eqnarray}
&&\frac{\theta_0}{2}=
\frac{r_{*}}{r_{+c}}\int_{z_{UV}}^{z_{*}}\frac{dz}{z\sqrt{f_0(z^{-2}-z_{*}^{-2})}}\label{theta-Tc}.
\end{eqnarray}
where
\begin{eqnarray}
T_c=\frac{1}{4\pi r_{+c}}\Big(2r_{+c}^2L^{-2}-\kappa^2\mu_c^2\Big)\label{Tc}
\end{eqnarray}
We go on to calculate the critical value of the horizon $r_{+}$ used by $\{T_c,\mu_c,\kappa^2\}$:
\begin{eqnarray}
\frac{r_{+c}}{L}=\pi \,T_{{c}}L+\frac{1}{2}\,\sqrt {4\,{\pi }^{2}{T_{{c}}}^{2}{L}^{2}+2\,{
\kappa}^{2}{\mu_{{c}}}^{2}}.
\end{eqnarray}
Parametric estimation of the (\ref{s-Tc},\ref{theta-Tc})  using
polynomials is needed. We apply series method to estimation of
$\frac{1}{\sqrt{f(z)}}$ in (\ref{s-Tc},\ref{theta-Tc}). Expansion of
the $\frac{1}{\sqrt{f(z)}}$ as follows :
\begin{eqnarray}
&&\frac{1}{\sqrt{f_0}}=\Sigma_{n=0}^{\infty}b_n\frac{(\log z)^n}{(z^{-2}-1)^{n+1/2}},\\&&
b_n=\frac{(\kappa \mu_c)^{2n}(-1)^n(1/2)_n\Big(\frac{L}{r_{+c}}\Big)^{2n+1}}{n!}
\end{eqnarray}
allows us  to expand into the series the (\ref{s-Tc},\ref{theta-Tc}). We first evaluate a value of integral $I^n_a$ for use in the  (\ref{s-Tc},\ref{theta-Tc}) :
\begin{eqnarray}
&&I^n_a\equiv\int_{z_{UV}}^{z_{*}}\frac{(\log z)^ndz}{z^a\sqrt{z^{-2}-z_{*}^{-2}}(z^{-2}-1)^{n+1/2}} ,\\&&\nonumber \mbox{for}  \ \  a=1,3
\end{eqnarray}
The aim is to evaluate the integral of $I^n_a$  as a series tool for interval $0\lesssim z_{UV}<z<z_{*}\lesssim 1$:
\begin{eqnarray}
&&I^n_a=\Sigma_{\alpha,\beta,\gamma=0}^{\infty}\frac{(1/2)_{\alpha}(n+1/2)_{\beta}}{\alpha!\beta!\gamma!(\gamma+n)}\\&&\nonumber\times \Big(2(\alpha+\beta+n)-a+3\Big)^{\gamma}\Big[(\log z_{*})^{\gamma+n}-(\log z_{UV})^{\gamma+n}\Big]
\end{eqnarray}
here $(a)_n\equiv\frac{(a+n-1)!}{n!(a-1)!}$ is  the Pochhammer symbol \cite{book}.
We need to carefully evaluate (\ref{s-Tc},\ref{theta-Tc}) with $I^n_a$:
\begin{eqnarray}
&&\frac{\theta_0}{2}=\frac{r_{*}}{r_{+c}}\Sigma_{n=0}^{\infty}b_n I_1^n,\\&&
s_{\tilde{A}}=2r_{+c}r_{*}\Sigma_{n=0}^{\infty}b_n I_3^n.
\end{eqnarray}
Indeed, the functions
$\theta'\equiv\frac{\theta_0}{2},s'\equiv\frac{s_{\tilde{A}}}{r_{+c}^2}$
have the simple forms in their bi-parametric
$\Big(\frac{T}{T_c},\frac{\mu}{\mu_c}\Big)$ list:
\begin{eqnarray}
&&s'=\frac{2\frac{T}{T_c}+\sqrt{4(\frac{T}{T_c})^2+\frac{2\zeta^2}{T_c^2\pi^2 L^2}(\frac{\mu}{\mu_c})^2}}{1+\frac{1}{2}\sqrt{4+\frac{2\zeta^2}{(\pi L T_c)^2}}}\Sigma_{n=0}^{\infty}B_n I_{3}^n\label{s'}\\&&
\theta'=\frac{\frac{T}{T_c}+\frac{1}{2}\sqrt{4(\frac{T}{T_c})^2+\frac{2\zeta^2}{T_c^2\pi^2 L^2}(\frac{\mu}{\mu_c})^2}}{1+\frac{1}{2}\sqrt{4+\frac{2\zeta^2}{(\pi L T_c)^2}}}\Sigma_{n=0}^{\infty}B_n I_{1}^n\label{theta'}
\end{eqnarray}
as they may  have the  parameters $\zeta=(\kappa \mu_c)=0.005,L\equiv 1$ and
\begin{eqnarray}
&&B_n=\frac{\zeta^n(-1)^n(1/2)_n}{n!}\\ \nonumber&&\times\Big(\pi \,T_{{c}}L+\frac{1}{2}\,\sqrt {4\,{\pi }^{2}{T_{{c}}}^{2}{L}^{2}+2\zeta^2}\Big)^{-(2n+1)}.
\end{eqnarray}


\par

\begin{figure}[htb]
\epsfxsize=6cm \epsfbox{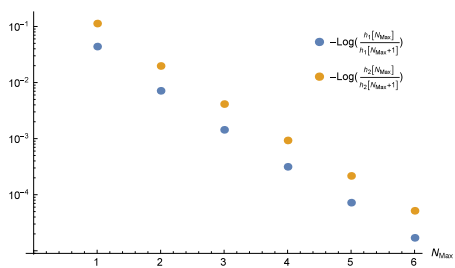} \caption{ The dependence of $h_1$ and $h_2$ functions on the upper limit of summation $n=N_{Max}$ in equations (\ref{theta'}) and (\ref{s'}). As one can see, the logarithm difference between $n=5$ and $n=6$ summations is less than $10^{-4}$. Therefore, we choose the $n=6$ as a series truncation in our calculations. This graph is for $T_c=0.01$.}\label{figseri}
\end{figure}

For numerical calculations we set the $T_c=0.01$.
If in equations  (\ref{s'}) and (\ref{theta'}), we define the $h_1(n)=\Sigma_{m=0}^{n}B_m I_{3}^m$ and $h_2(n)=\Sigma_{m=0}^{n}B_m I_{1}^m$ then we can show the $h_1$ and $h_2$ as a function of $n$ in figure (\ref{figseri}).
We choose the $n=6$ as high value for $n$, because we can omit the relative error arising from series truncation.
\par
\begin{figure}[htb]
\epsfxsize=6cm \epsfbox{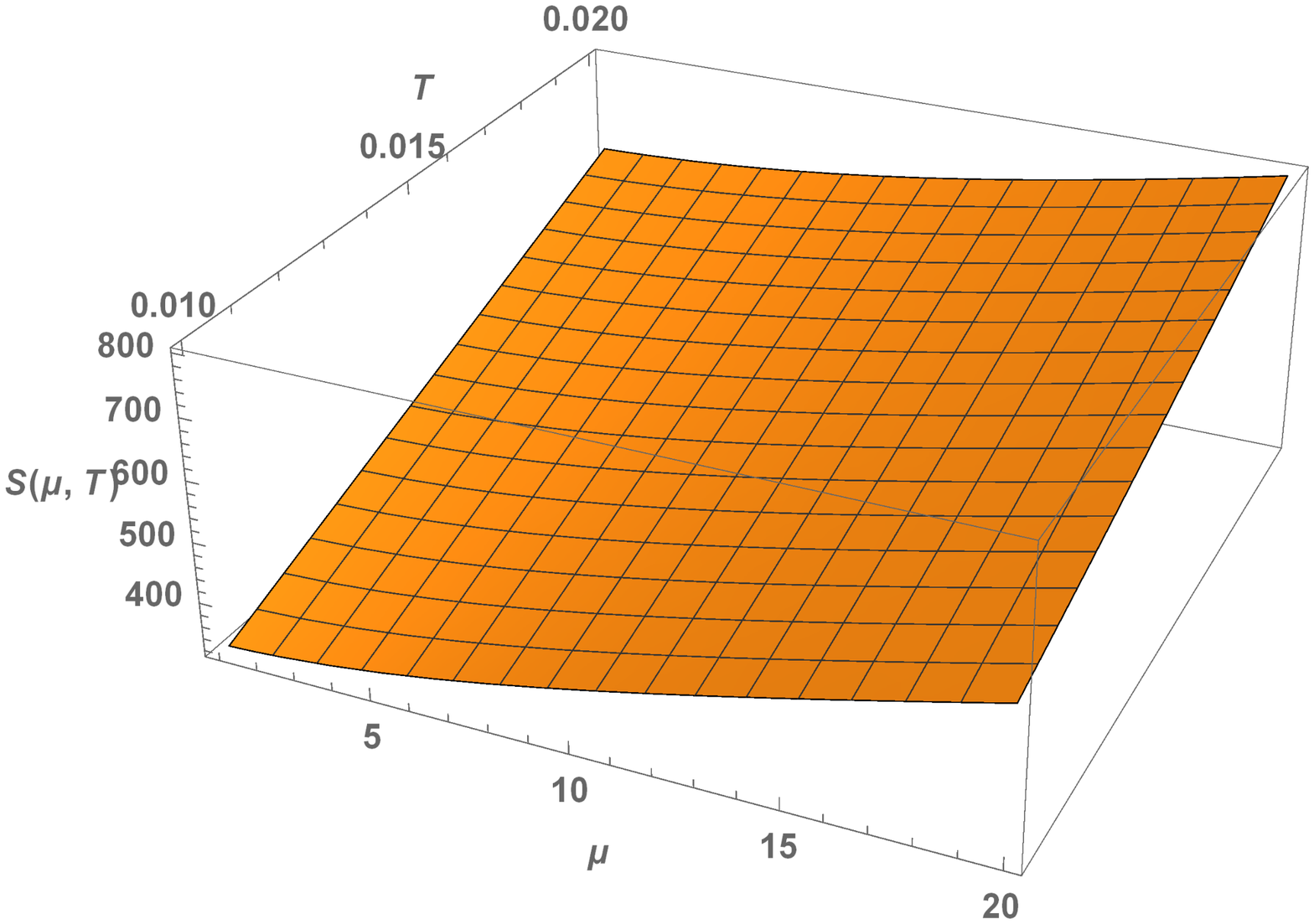} \caption{Plot of the surface
(\ref{s'}) versus $\mu,T$.  It shows that $s'$ is a
monotonic-increasing function.  It always increasing or remaining
constant, and never decreasing. It produces a regular phase of
matter for $T>T_c$. Regular attendance at these non superconducting
phase has proved numerically. Boundary conditions and regular tiny
backreactions $\zeta$ will help to keep normal phase for longer.
Normal phase increasing the entropy (\ref{s'}),  increases the
hardenability of superconductivity.}\label{fig1}
\end{figure}

An example (\ref{s'}) plot of reduced HEE  in a system is shown in
figure (\ref{fig1}). Numeric analysis showed a significant smooth
relationship between increasing proportions of $\mu,T$ and increased
$s'$ in the this phase. Seeing an increasing HEE for system, it
decided to become a  normal conductor than just a superconductor.
Increasing temperature to reduce HEE adds to the system a
criticality, thus slowing the superconducting.

\begin{figure}[htb]
\epsfxsize=6cm \epsfbox{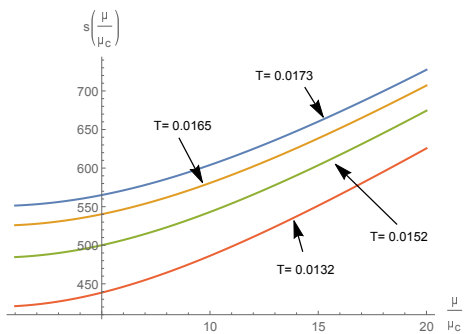} \caption{Plot of the of the
isothermal HEE  (\ref{s'}) for various values of  $T$. For positive
values of the chemical potential,  $s'$ is a monotonic-increasing
function, always increasing , and never decreasing. The " fixed "
chemical potential  will give a lower value of (\ref{s'}) for lower
temperatures. At fixed temperature, in an isothermal graph, when
$\frac{\mu}{\mu_c}$  increases, the associated HEE entropy is also a
monotonic-increasing function of $\frac{\mu}{\mu_c}$ . Of course,
attending at least one lower temperature regime  $T<T_c$ is almost
compulsory for superconductivity. It was always realistic to expect
that superconducting phase could be in being by lower values of
$\frac{\mu}{\mu_c}$ in isothermal regime. We don't detect any
local maxima for $\frac{\mu}{\mu_c}$. Consequently no
"confinement/deconfinemnet" phase transition point  exists in our
$(1+1)D$ dual theory. }\label{fig2}
\end{figure}

We plot isothermal curves of (\ref{s'}) for various values of  $T$
in figure (\ref{fig2}). Attending at least one lower temperature
regime  $T<T_c$ is almost compulsory for superconductivity. We
don't detect any local maxima for $\frac{\mu}{\mu_c}$. Consequently
no "confinement/deconfinemnet" phase transition point  exists in our
$1+1$ dual theory.

\begin{figure}[htb]
\epsfxsize=6cm \epsfbox{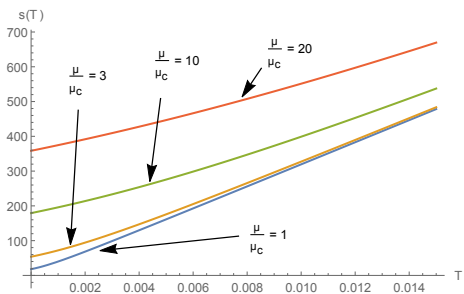}\caption{Plot of (\ref{s'}) as function of $T$ for various values of  $\frac{\mu}{\mu_c}$. At fixed  $T$ , we may increase  the $s'(T)$ simply by  increasing the $\frac{\mu}{\mu_c}$, or by increasing  the number of Cooper  (BCS) pair. Formation of the Cooper pairs decreses the extra dof of system. It is important to exclude $\frac{\mu}{\mu_c}=1$ for system. We should entirely exclude phase transition critical point $\mu=\mu_c$  in our general study of HEE (\ref{s'}) against the cases $\frac{\mu}{\mu_c}>1$.  A somewhat amazing statement considering the HEE attempt to exclude the superconducting phase from the normal phase. This keyword can be used to exclude part of the  criticality by entropy expression.  At fixed $\frac{\mu}{\mu_c}$,  one may increase  the $s'(T)$ simply by  increasing the $T>T_c$. Furthermore we observe that the slope of the HEE with respect to the temperature
$\frac{dS}{dT}$ decreases, thanks to the emergence extra dof in low temperature system. }\label{fig3}
\end{figure}

For fixed relative chemical potential $\frac{\mu}{\mu_c}$, we plot
(\ref{s'}) as function of $T$ in figure (\ref{fig3}). We observe
that at fixed $\frac{\mu}{\mu_c}$,  one may increase  the $s'(T)$
simply by  increasing the $T$. Furthermore, we see that the
slope of the HEE with respect to the temperature $\frac{dS}{dT}$
decreases as the relative chemical potential
$\frac{\mu}{\mu_c}\neq1$ decreases. We understand this through the
fact that, in low temperature and $\frac{\mu}{\mu_c}\neq1$, more
degree of freedoms (dof) will condense. An emergent of new extra dof
at low temperature is happening. 

\begin{figure}[htb]
\epsfxsize=6cm \epsfbox{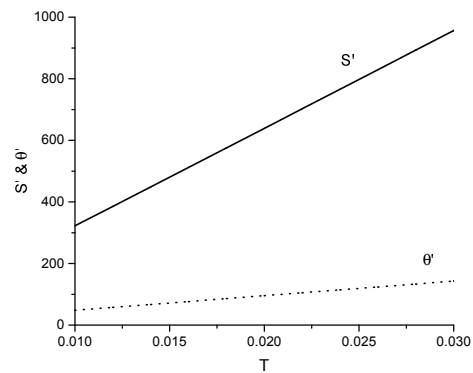} \caption{The entanglement entropy  (\ref{s'}) and the angle $\theta'$ as a function of $T$ for fixed $\frac{\mu}{\mu_c}$. }\label{fig4}
\end{figure}

Figure (\ref{fig4}) shows typical behaviors of
(\ref{s'},\ref{theta'}) versus temperature $T$ for fixed $T_c$.
Both are always increasing with respect to the temperature $T$, and
never decreasing. This type of monotonic-increasing behavior with
$T$ depends on thermodynamically stability condition, in which the
heat capacity at constant size must be positive. These are
relatively low temperature, holographic superconductors which
contain a prepared HEE which can linearly be described for system.
It has been suggested that where there is low temperature phase  may
be able to keep superconductivity with increased  reduced entropy
(\ref{s'}) rises.

\begin{figure}[htb]
\epsfxsize=6cm \epsfbox{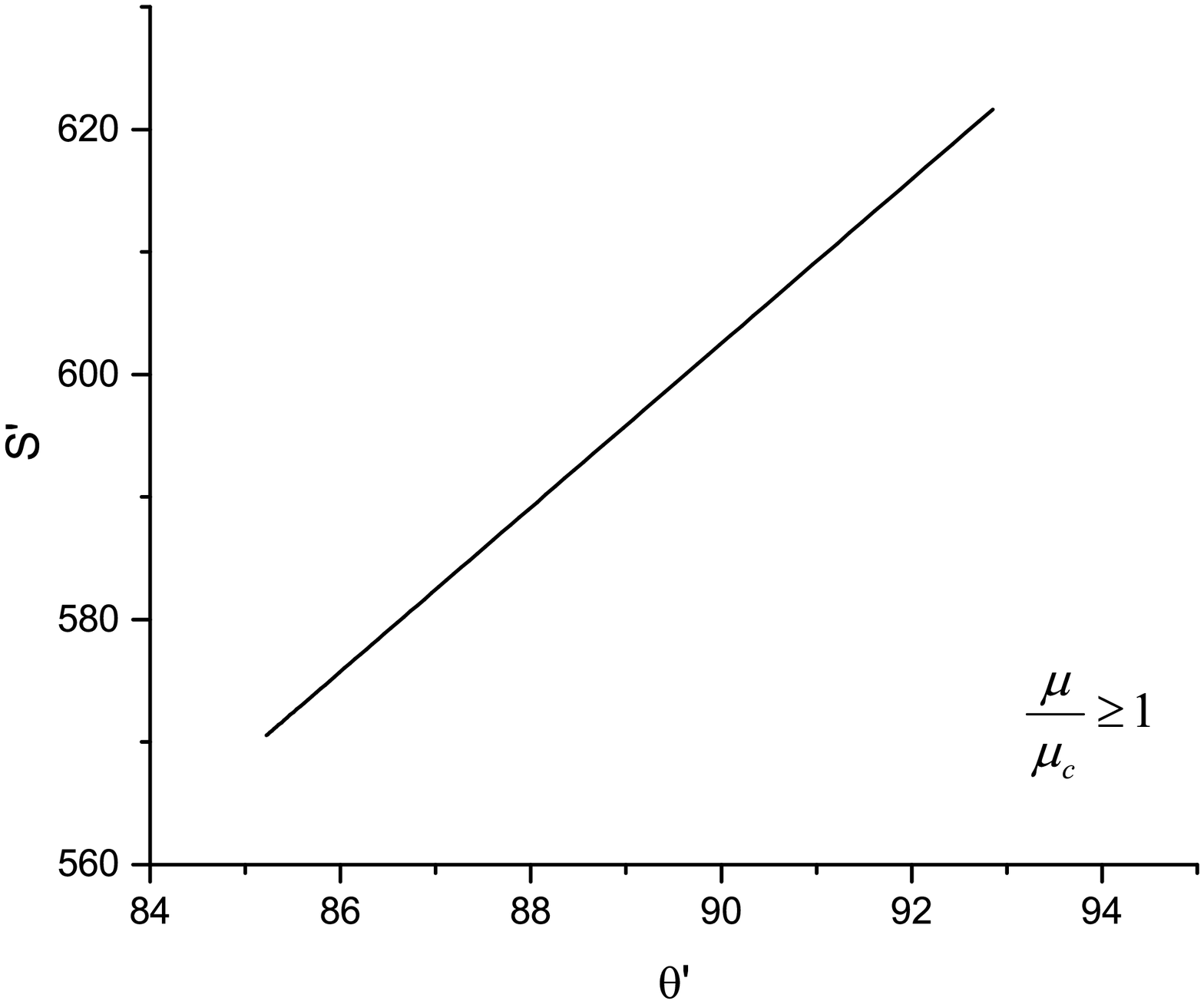} \caption{The entanglement
entropy (\ref{s'}) as a function of belt angle at fixed temperature
in the case of $\frac{\mu}{\mu_c}\geq1$. We observe that  HEE
(\ref{s'}) is dominated by the connected minimal surface. The wider
angle (\ref{theta'}) corresponds to a larger surface holographic
surface. We see that HEE (\ref{s'}) changes linearly with $\theta'$.
A simple computational reason "why the $s'$ can dominate on
$\theta'$", is that  the main contribution (\ref{s'},\ref{theta'})
comes from the region $r\sim r_{*}\sim r_{+}$ or $z\sim z_{*}$.
Furthermore, there is no critical belt angle $\theta'_{c}$ in which
we can label the "confinement/deconfinement" transition point to it.
The main reason is that the HEE is an extensive function of belt
angle, $s'(k\theta')=ks'(\theta')$. }\label{fig5}
\end{figure}

Figure (\ref{fig5}) shows a linearly-dependent of  reduced HEE
(\ref{s'}) versus angle (\ref{theta'}). The physical reason is
that in small values of belt angle (small sizes) the system emerges
new extra dof. A simple computational reason "why the $s'$ can
dominate on $\theta'$", is that  the main contribution
(\ref{s'},\ref{theta'}) comes from the region $r\sim r_{*}\sim
r_{+}$ or $z\sim z_{*}$. A better more simple reason can be
understood through the first law of thermodynamic for entanglement
entropy. As we know, HEE behaves like a conventional entropy and it
obeys the first law of thermodynamic
\cite{Bhattacharya:2012mi},\cite{Momeni:2015vka}. If we consider
$\theta'$ as the length scale of the system, then
$\frac{ds'}{d\theta'}$ is proportional to the entangled pressure
$P_{E}=T_{E}\frac{ds'}{d\theta'}$  at fixed temperature in the case
of $\frac{\mu}{\mu_c}>1$. A constant slope $\frac{ds'}{d\theta'}$
gives us a uniform entangled pressure $P_{E}$. From the Maxwell’s
relations we know that
$\Big(\frac{ds'}{d\theta'}\Big)_{T}=\Big(\frac{dP}{dT}\Big)_{\theta'}$.
It means that at fixed $T$, there is  a uniform entropic gradient of
HEE $\Big(\frac{ds'}{d\theta'}\Big)$. Consequently we obtain a
uniform  gradient of pressure $\Big(\frac{dP}{dT}\Big)$ at fixed
belt angle. A constant entropic force is emerged
\cite{Bhattacharya:2012mi}. Another physical reason is that $s'$
must be an extensive function of the "volume" or "size" of the
entangled system, namely $\theta'$. Within the statistical mechanics
there are extensive parameters like size, number of particles  and
thermodynamical functions like entropy.
 If we increase the size of the entangled system, here $\theta'\to k\theta'$, then the HEE $s'$ must also increases.
 It means that $s'$ must be a homogenous function of size. In this
case, $s'$ is found to be homogenous  of first order, i.e.
$s'(k\theta')=ks'(\theta')$. Consequently $s'\sim\theta'$ changes
linearly with $\theta'$.

\begin{figure}[htb]
\epsfxsize=6cm \epsfbox{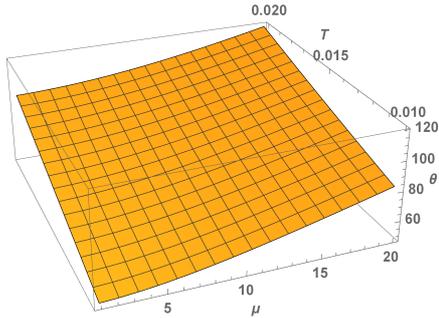} \caption{3D plot of $\theta'$ as a function of $\mu$ and $T$.  It shows that $\theta'$ is a monotonic-decreasing function.}\label{fig6}
\end{figure}

Figure (\ref{fig6}) shows that there are low-impact angle
(\ref{theta'}) designed specifically for low temperature and
chemical potential. Furthermore, $\theta'$ is a monotonic-decreasing
function of $\mu,T$.

\par
{\it  HEE in the presence of scalar field $\phi(z)\neq0$ at  $T \lesssim T_c$}:
During the critical phase transition, $\epsilon\equiv <\mathcal{O}_{\pm}>$   is sufficiently tiny  to expand functions by the following series forms:
\begin{eqnarray}
&&\phi=\Sigma_{k=1}^{\infty}\epsilon^k\phi_{k},\ \ A=\Sigma_{k=0}^{\infty}\epsilon^{2k}A_{2k},\\&&
f=\Sigma_{k=0}^{\infty}\epsilon^{2k}f_{2k},\ \ \beta=\Sigma_{k=1}^{\infty}\epsilon^{2k}\beta_{2k}.
\end{eqnarray}
When we turn-on the condensate, $\phi(z)\neq0$, as we expect,
analytical expression for HEE is much more  harder. Specially at the
criticality,  $T \lesssim T_c$, because  scalar field $\phi(z)$ and
Maxwell field $A(z)$ backreacted on the metric functions
$f(z),\beta(z)$. Generally speaking, one cannot solve  field
equations given in (\ref{phiz}-\ref{fz}) and find the analytic form
of  $f(z)$ in a closed form. However, the approximate solutions for
(\ref{phiz}-\ref{fz}) will be possible. We start by the following
solutions:
\begin{eqnarray}
&&\phi(z)=\epsilon \phi_1,A(z)=A_0+\epsilon^2 A_2,\label{aprox1}\\&&
\beta(z)=\epsilon^2\beta_2,f(z)=f_0+\epsilon^2f_2\label{aprox2}.
\end{eqnarray}
where $\epsilon\equiv<\mathcal{O}_{\pm}>$. Analytical solutions obtained by substituting (\ref{aprox1},\ref{aprox2}) into the field equations (\ref{phiz}-\ref{fz}):
\begin{eqnarray}
&&f_2=\Big[-2\,{\kappa}^{
2}\mu_{{c}}B+{\kappa}^{2}{B}^{2}+\,{\frac {4{r_{{+c}}}^{2}}{{L}^{2}}}\Big](1-z)\\&&\nonumber+\mathcal{O}((1-z)^2) ,
\ \mbox{as } T \lesssim T_c.
\end{eqnarray}
Which can be used according to the approximate solutions of the
fields:
\begin{eqnarray}
&&\phi_1=\mu(1-z),\ \ A_2=B(1-z),\ \ \beta_2''(1)= 0.
\end{eqnarray}
we can approximate the HEE  and belt angle  by putting a metric function  through a carefully defined integrals:
\begin{eqnarray}
&&s_{\tilde{A}}=2r_{+c}r_{*}\int_{z_{UV}}^{z_{*}}\frac{dz}{z^3\sqrt{f_0+\epsilon^2 f_2}\sqrt{z^{-2}-z_{*}^{-2}}}\\&&
\frac{\theta_0}{2}=
\frac{r_{*}}{r_{+c}}\int_{z_{UV}}^{z_{*}}\frac{dz}{z\sqrt{(f_0+\epsilon^2 f_2)(z^{-2}-z_{*}^{-2})}}.
\end{eqnarray}
where $\epsilon\sim \sqrt{\mu-\mu_c}\sim\sqrt{1-\frac{T}{T_c}}\ll1$.\par
Expansion of the $\frac{1}{\sqrt{f_0+\epsilon^2 f_2}}$ as follows :
\begin{eqnarray}
\frac{1}{\sqrt{f_0+\epsilon^2 f_2}}=\frac{1}{\sqrt{f_0}}\Big(1-\frac{1}{2}\epsilon^2\frac{f_2}{f_0}\Big).
\end{eqnarray}
we obtain:

\begin{eqnarray}
&&\frac{\theta_0}{2}=\frac{r_{*}}{r_{+c}}\Sigma_{n=0}^{\infty}b_n \Big(I_1^n-\frac{1}{2}\epsilon^2\tilde{I}_1^n\Big),\\&&
s_{\tilde{A}}=2r_{+c}r_{*}\Sigma_{n=0}^{\infty}b_n \Big(I_3^n-\frac{1}{2}\epsilon^2\tilde{I}_3^n\Big).
\end{eqnarray}
Where

\begin{eqnarray}
&&\tilde{I}^n_a\equiv\int_{z_{UV}}^{z_{*}}
\frac{f_2}{f_0}\frac{(\log z)^ndz}{z^a\sqrt{z^{-2}-z_{*}^{-2}}(z^{-2}-1)^{n+1/2}} ,\\&&\nonumber \mbox{for}  \ \  a=1,3
\end{eqnarray}
We rewrite them in terms of $\Big(\frac{T}{T_c},\frac{\mu}{\mu_c}\Big)$ as the following:
\begin{eqnarray}
&&\label{s'2}s'=\frac{\frac{2T}{T_c}+\sqrt{4(\frac{T}{T_c})^2+\frac{2\zeta^2}{T_c^2\pi^2 L^2}(\frac{\mu}{\mu_c})^2}}{1+\frac{1}{2}\sqrt{4+\frac{2\zeta^2}{(\pi L T_c)^2}}}\\&&\nonumber\times\Sigma_{n=0}^{\infty}B_n \Big(I_3^n-\epsilon_0^2\frac{1-\frac{T}{T_c}}{2}\tilde{I}_3^n\Big)\\&&
\label{theta'2}\theta'=\frac{\frac{T}{T_c}+\frac{1}{2}\sqrt{4(\frac{T}{T_c})^2+\frac{2\zeta^2}{T_c^2\pi^2 L^2}(\frac{\mu}{\mu_c})^2}}{1+\frac{1}{2}\sqrt{4+\frac{2\zeta^2}{(\pi L T_c)^2}}}\\&&\nonumber\times\Sigma_{n=0}^{\infty}B_n \Big(I_1^n-\epsilon_0^2\frac{1-\frac{T}{T_c}}{2}\tilde{I}_1^n\Big)
\end{eqnarray}
Here $\epsilon_0\ll1$ is a numeric. The second negative term,  seems
obviously compatible with a superconductor phase in the presence of
the scalar field. \emph{By decreasing the amount of entropy produced
in the superconductor phase,} system alters the phase of
conductivity.\par
For numerical calculations in  (\ref{s'2}) and (\ref{theta'2}) we must estimate the numerical errors. As figure (\ref{figseri}), for low temperature region, in this case we have:
\begin{eqnarray}
h_{1(2)}(N_{max})= \sum_{n=0}^{N_{Max}} b_n (I_{3(1)}^n-\frac{\epsilon_0^2}{2} (1-\frac{T}{T_c})\tilde{I}_{3(1)}^n).
\end{eqnarray}

\begin{figure}[htb]
\epsfxsize=6cm \epsfbox{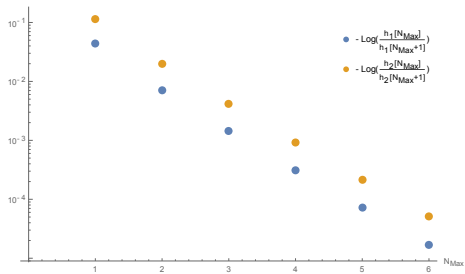} \caption{ The graph represent zero temperature case with $T_c=0.01$, because it has the maximum difference with figure (\ref{figseri}). Similar to figure (\ref{figseri}), we choose $N_{Max}=6$.}\label{figseri2}
\end{figure}

\par
We plot (\ref{s'2})  vs. (\ref{theta'2}). We adjust data as
$\epsilon_0^2 = 0.05,\kappa \mu_c=0.005$. The critical temperature
was obtained as $T_c=0.2$. The system evolves from normal phase
$T>T_c$ to the superconductor phase $T \lesssim T_c$ for $T\approx
0.0179, 0.0173, 0.0165, 0.0152, 0.0132$. The wider angle
(\ref{theta'2}) corresponds to a larger surface holographic surface.
We see that HEE (\ref{s'2}) changes linearly with $\theta'$. We
observe that the slope of the HEE with respect to the belt angle
$\frac{ds'}{d\theta'}$ remains constant. Like the non superconductor
phase, here is no critical belt angle $\theta'_{c}$ in which we can
label the "confinement/deconfinement" transition point to it. The
main reason is that the HEE is an extensive, homogenous (first
order)  function of belt angle , $s'(k\theta')=ks'(\theta')$.

\begin{figure}[htb]
\epsfxsize=6cm \epsfbox{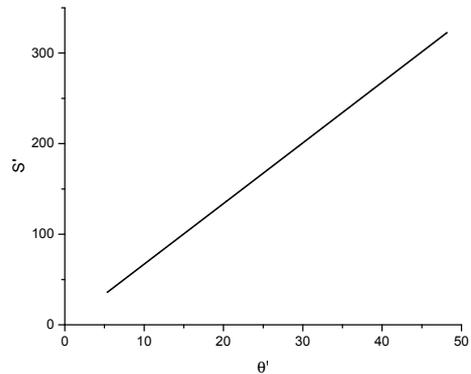} \caption{Plot of $s'$
(\ref{s'2}) as a function of $\theta'$ (\ref{theta'2}) for $T
\lesssim T_c$. We adjust data as  $\epsilon_0^2 = 0.05$.   The
critical temperature was obtained as $T_c=0.2$. The system evolves
from normal phase $T>T_c$ to the superconductor phase $T \lesssim
T_c$. The wider angle (\ref{theta'2}) corresponds to a larger
holographic surface. We see that HEE (\ref{s'2}) changes linearly
with $\theta'$. }\label{fig7}
\end{figure}
\par
\begin{figure}[htb]
\epsfxsize=6cm \epsfbox{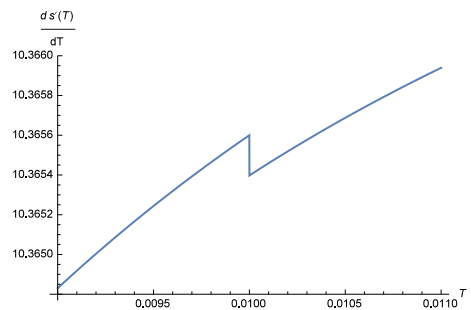} \caption{Discontinuity in $\frac{ds'}{dT}$ near critical point $T_c=0.01$. We scaled the entropy in $\log$-scale form. }\label{figdis}
\end{figure}

{\it  Phase transition at critical point }: A numerical study
of (\ref{s'}),(\ref{s'2}) shows that these solutions were smooth,
and that their behaviors went most smoothly when the phase
transition held the same mechanism as the usual. But after the
system proceeded more smoothly, and the temperature of the system
$T$ regained his critical value $T_c$ in the system, if the
temperature alters as the $T\simeq T_c$, a discontinuity occurs in
the $\frac{d s'}{dT}$  when the phase is changed, and a  first order
phase transition may be introduced into the system. The difference
between $\frac{d s'}{dT}$ for $T > T_c$ (eq.(\ref{s'} )) and $T< T_c
$  (eq.(\ref{s'2} )) at $T=T_c $ is plotted in figure (\ref{figdis})
for a log-scaled entropy. The graph is obtained by smoothly
connecting two graphs of $s'(T)$ in the normal phase $T>T_c$ i.e.
the figure (\ref{fig4}) and the one in the superconductor phase
based on the formula given in (\ref{s'2}). When we scaled the
entropy in $\log$ scale, we observe  a first order discontinuity in
$\frac{d s'}{dT}$ at the critical point $T=T_c$. Indeed, at the
critical point $\lim \frac{d s'}{dT}|_{T\to T_c}=\infty$ and $
\frac{d s'}{dT}|_{T> T_c}- \frac{d s'}{dT}|_{T< T_c}\simeq  \sum
_{n=0}^{\infty } B(n)\tilde{I}_3^n $. We observe the  first order
phase transitions from the behavior of the entanglement entropy
$s'(T)$ at the critical point $T=T_c$.  These types of first order
phase transitions have been observed recently in literature
\cite{Li:2013rhw}. We conclude that the HEE  is indeed a good probe
to phase transition in lower dimensional holographic
superconductors. Furthermore, it implies that the HEE can indicate
not only the occurrence of the phase transition, but also we can
learn about  the order of the phase transition from it.

\par
{\it Summary:} The aim of this letter was to investigate the effect
of superconductor critical phase transition in 2D models of
holographic superconductors  on   holographic entanglement entropy.
 We investigate analytical aspects of the domain wall approximation and
scalar condensate of the transition phases. Using the domain wall
auxiliary asymptotic boundary conditions,  as have been used before
we can investigate the evolution of the  holographic entanglement
entropy for this superconductor model. To calculate the HEE in the
critical phase first the interval is divided by the cutoffs in the
UV and IR domains. Then we have to resort the calculations of the
holographic entanglement entropy in the presence of scalar field. It
can be computed analytically in terms of the series functions of
$\mu,T$. After the normal phase $T>T_c$, the superconductor phase $T
\lesssim T_c$  for the equations has been derived, the extent to
which the equations may be solved analytically is covered. In case
we cannot calculate minimal surface integrals analytically it offers
the possibility to proceed with a numerical evaluation of the
corresponding terms. We proceeded to investigate why the HEE
increase with  temperature and belt angle in the backreacted and
normal $AdS_3$ background.  Both are always increasing with respect
to the temperature $T$ and belt angle $\theta'$, and never
decreasing. This type of monotonic-increasing behavior with $T$
depends on thermodynamical stability condition, in which the heat
capacity at constant size must be positive. In the case of
$\theta'$, there is no critical belt angle $\theta'_{c}$ in which we
can label the "confinement/deconfinement" transition point to it.
The main reason is that the HEE is an extensive, homogenous (first
order)  function of belt angle, $s'(k\theta')=ks'(\theta')$.
 We observe the first order  phase transitions from the behavior of
the entanglement entropy  $s'(T)$ at the critical point $T=T_c$. We
conclude that the wider belt angle corresponds to a larger surface
holographic surface. Hopefully, the results of this study would come
out until we could explore the roles of backreactions and scalar
condensation on holographic entanglement entropy.\par {\it Acknowledgments}
 We are  indebted to the referee for pointing out
the important comments and making suggestions for improvement of
this work.


\begin{thebibliography}{90}
\bibitem{Tameem Albash}
We are indebted to Tameem Albash for pointing out this comment.
\end{thebibliography}
\end{document}